\documentclass[11pt,a4paper]{article}
 
\usepackage{amsmath,amsthm,amssymb}
\usepackage{graphics,graphicx}
\usepackage{epsfig}
\usepackage{multicol}
\usepackage{color}
\makeatletter
\@addtoreset{equation}{section}
\makeatother
\renewcommand{\theequation}{\thesection.\arabic{equation}}

\newcommand{\n}{\noindent}

\begin{document}

\begin{flushright}
{QMUL-PH-09-26}
\end{flushright}
\begin{center}
\LARGE{\bf Non-Topological Cycloops}
\end{center}

\begin{center}
\large{\bf Matthew Lake} ${}^{a, b, }$\footnote{m.lake@qmul.ac.uk} \large
{\bf Steven Thomas} ${}^{a, }$\footnote{s.thomas@qmul.ac.uk}  \large{\bf and John Ward} ${}^{c, }$\footnote{jwa@uvic.ca}
\end{center}
\begin{center}
\emph{ ${}^a$ Center for Research in String Theory, Queen Mary University of London \\ Mile End Road, London E1 4NS, UK \\}
\vspace{0.1cm}
\emph{ ${}^{b}$ Astronomy Unit, School of Mathematical Sciences, Queen Mary University of London \\ Mile End Road, London E1 4NS, UK \\}
\vspace{0.1cm}
\emph{${}^{c}$ Department of Physics and Astronomy, University of Victoria, Victoria, BC \\ V8P 1A1, Canada }
\end{center}

%Opening%%%%%%%%%%%%%%%%%%%%%%%%%%%%%%%%%%%%%%%%%%%%%%%%%%%%%%%%%%%%%%%%%
%%%%%%%%%%%%%%%%%%%%%%
%Abstract%%%%%%%%%%%%%%%%%%%%%%%%%%%%%%%%%%%%%%%%%%%%%%%%%%%%%%%%%%%%%%%%
%%%%%%%%%%%%%%%%%%%%%%

\begin{abstract}
We propose a mechanism for the creation of cosmic string loops with dynamically stabilised windings in the internal space. 
Assuming a velocity correlations regime in the post-inflationary epoch, such windings are seen to arise naturally in string networks 
prior to loop formation. The angular momentum of the string in the compact space may then be sufficient to ensure that the 
windings remain stable \emph{after} the loop chops off from the network, even if the internal manifold is simply connected. 
For concreteness we embed our model in the Klebanov-Strassler geometry, which provides a natural mechanism for brane inflation, 
as well a being one of the best understood compactification schemes in type IIB string theory. 
We see that the interaction of angular momentum with the string tension causes the loop to oscillate between phases of expansion and contraction. 
This, in principle, should give rise to a distinct gravitational wave signature, the future detection of which could provide indirect evidence for the 
existence of extra dimensions. 
\end{abstract}

%Section1%%%%%%%%%%%%%%%%%%%%%%%%%%%%%%%%%%%%%%%%%%%%%%%%%%%%%%%%%%%%%%%%
%%%%%%%%%%%%%%%%%%%%%%%

\section{Introduction}
The existence of string loops with dynamically stabilised winding in a compact space was first demonstrated by Iglesias and Blanco-Pillado 
\cite{BlancoPillado:2005dx}, using the Klebanov-Strassler geometry \cite{Klebanov:2000hb}. 
They considered strings at the tip of the throat, with geodesic wrappings in the $S^3$ 
which regularises the conifold singularity. Although they derived a lower bound for the angular momentum of a loop with a given number of windings - 
below which the windings became unstable - the result must remain of purely theoretical interest to cosmology as long as a 
specific mechanism for winding formation (and hence for the formation of the string angular momentum) is not considered. 
The purpose of this paper is to propose such a mechanism, which leads to the formation of string configurations such as those investigated in 
\cite{BlancoPillado:2005dx} and to investigate the resulting string dynamics with specific reference to their cosmologically observable consequences.

\n
Assuming that a velocity correlations regime in the post-inflationary epoch as in \cite{Avgoustidis:2005vm} leads naturally to the formation of geodesic 
windings we show that the winding number ($n$), total energy ($E$) and angular momentum ($l$) of the string are 
specified precisely by the model parameters. That is, by the parameters which define the Klebanov-Strassler geometry (in this case specifically by the value of the warp factor $a_0$) 
and those which determine the scale of the string network ($\alpha$, $t_i$). Substituting for $l$ and $n$ in the bound referred to above 
then demonstrates the stability of these windings, at least under the assumption that $l$ remains approximately constant over small time scales 
after the moment of initial loop formation. By assuming also that the total energy of the string remains approximately constant 
(i.e. by neglecting the loss of $E$ and $l$ via gravitational wave emission over small time scales), 
we then determine the equation of motion for the four dimensional string radius $r(t)$, and solve to find a (generically) oscillating solution. 
Crucially we observe that the qualitative behaviour of the loop depends on the value of the warp factor $a_0$ with $a_0^2<1/2$ leading to an 
initial phase of expansion and $a_0^2>1/2$ leading to an initial phase of contraction. The fixed point solution $a_0^2 = 1/2$ is a static, non-oscillatory
solution. In both oscillatory modes (initially contracting \emph{and} initially expanding) we find that the period of the oscillation is inversely 
proportional to $a_0^2$, and proportional to the initial size of the loop $(\alpha t_i)$. 
Following \cite{Avgoustidis:2005vm} we refer to these objects as non-topological cycloops \footnote{The term 'cycloop' was first coined in \cite{Avgoustidis:2005vm} 
to refer to cosmic string loops with \emph{smooth} windings wrapping 
cycles in the internal space - as opposed, for example, to non-smooth, step-like windings which give rise to necklace configurations from a 
four dimensional perspective \cite{Matsuda:2006ju, Matsuda:2005ez, Lake:2009nq}. However in their original conception 
Avgoustidis and Shellard used it only to refer to windings which are topologically trapped. 
We therefore propose the term 'non-topological cycloops' to refer to string loops with dynamically 
stabilised smooth windings (in this case geodesics) around a simply connected compact manifold.}.

\n
The layout of the paper is then as follows: In Section 2 we briefly review the relevant Klebanov-Strassler background, 
focusing on the geometry of the conifold tip. In Section 3 we show how the assumption of a velocity correlations regime yields a 
dynamical model of winding formation after the end of inflation. Section 4 recaps the generic results of \cite{BlancoPillado:2005dx} 
which we then combine with the results of the previous section to give explicit expressions for the winding number, energy and angular momentum of 
the string. The equation of motion for the loop radius is then derived and solved in Section 5 and a brief summary of the main results 
together with a discussion of their cosmological implications and possible consequences for experimental observations 
is presented in section 6. Finally the two appendices at the end of this work deal with issues which arise within the text: 
Appendix I outlines the method of Eulerian substitution of the third kind which is used to integrate the 
differential equation involving $\dot{r}(t)$ and $r(t)$ derived in Section 5. 
Appendix II gives a detailed description of the Hopf fibration of the 3-sphere, which is introduced briefly in section 2 and used throughout the following analysis.   

%***************************************************************************************************************************************************
\section{The Klebanov-Strassler Geometry}
The Klebanov-Strassler geometry is the canonical example of a background which resolves a conifold singularity in type IIB string theory.
We refer the interested reader to the original paper \cite{Klebanov:2000hb} for a more thorough discussion. The essential point is that the conifold is the 
cone over an $S^2 \times S^3$ base space. When we deform the conifold, the $S^2$ shrinks to zero size, and the ten dimensional metric factorises
into the (warped) product of $\mathbb{R}_{1,3} \ltimes S^3$.
In canonical coordinates the effective metric of the Klebanov-Strassler geometry at the tip of the warped throat takes the following form;
\begin{equation} \label{eq:metric}
ds^2 = a_0^2 \eta_{\mu \nu}dx^{\mu}dx^{\nu} + R^2(d\psi^2 + \sin^2\psi(d\theta^2 + \sin^2\theta d\phi^2))  
\end{equation}
where $\eta_{\mu \nu}$ is the usual four-dimensional Minkowski metric, $a_0^2$ is the square of the warp factor such 
that $0 < a_0^2 < 1$ and $R$ is the radius of the three-sphere defined by,
\begin{equation} \label{eq:R^2}
R^2 = bMg_s\alpha'.
\end{equation}
Here $M$ is the number of units of flux wrapping the internal space, which is also equal to the number of fractional $D$3-branes at the bottom of the throat, 
$g_s$ is the string coupling, $l_s = \sqrt{\alpha'}$ is the fundamental length scale of the string 
and $b$ is a numerical constant of order one. 
The relation between the size of the $S^3$ and the warp factor induced by the back reaction of the fluxes is,
\begin{equation} \label{eq:a_0^2vsR^2}
a_0^2 \sim \frac{\tilde{\epsilon}^{-4/3}}{R^{2}}
\end{equation} 
where the constant $\tilde{\epsilon}^{-4/3}$ is the deformation parameter of the conifold.
Of course the relations (\ref{eq:R^2}) and (\ref{eq:a_0^2vsR^2}) remain true in any coordinate system, and we will find it 
convenient to use the Hopf fibration of the three-sphere when considering the formation of geodesic windings. 
These are the kind of windings we expect to form in the presence of velocity correlations 
which impart an initial (constant) angular momentum density to each point along the string. 
In Hopf coordinates the $S^3$ is described as a one-parameter family of flat two-tori (to which it is topologically equivalent) 
\cite{GQS:2006} and the canonical metric (\ref{eq:metric}) reduces to a much simpler form, 
\begin{equation} \label{eq:metric2}
ds^2 = -a_0^2 \eta_{\mu \nu}dx^{\mu}dx^{\nu} + R^2(d\psi^2 + d\theta^2 + \cos\theta  d\phi^2).
\end{equation}
The Killing vectors also adopt a simple form, and are always parallel to the unit vectors in the $\psi$, $\theta$ and $\phi$-directions. 
Fixing the value of the $\theta$-coordinate such that $\theta=\theta_0$ then selects a flat $T^2$ sub-manifold and windings which follow the 
Killing-directions in this manifold are necessarily geodesic in the full $S^3$. 
However the choice of gauge in this respect is somewhat arbitrary, as we are free to choose $\theta_0=0$. 
This simplifies both the resulting metric and the Killing vectors of the $T^2$, the latter now being \emph{identical} to the unit vectors in the 
remaining $\psi$ and $\phi$-directions. Although it may be shown explicitly that the Lagrangian density $\mathcal{L}$ for a string loop with geodesic windings 
in the $S^3$ is $\sigma$-independent in \emph{any} coordinate system, the simple form of the Killing vectors in Hopf coordinates 
allows us to more easily calculate $L = \int d\sigma \mathcal{L} = 2\pi \mathcal{L}$. 
A thorough treatment of the Hopf fibration of the three-sphere, and of the description of geodesic windings in both canonical and 
Hopf coordinates is given in Appendix II, along with a coordinate-independent geometric analysis. 
%*****************************************************************************************************************************************
\section{A dynamical model of winding formation}
We now proceed to construct our dynamical model of winding formation: 
In the velocity correlations regime the number of windings per loop - in a loop of size $r(t_i) = \alpha t_i$ - is \cite{Avgoustidis:2005vm},
\begin{equation} \label{eq:n1}
n(t_i) \sim \frac{\omega_l \alpha t_i}{R}
\end{equation}
where $\omega_l$ is the fraction of the total string length which lies in the extra dimensions and is defined via \cite{Lake:2009nq},
\begin{equation} \label{eq:omega_l_1}
\omega_l \sim \frac{nR}{\sqrt{a_0^2 r^2 + n^2 R^2}}.
\end{equation}
Substituting this back into the expression above gives a unique physical physical solution, 
%($n(t_i) > 0 \forall t_i > 0$) for $a_0^2 < 1$
\begin{equation} \label{eq:n2}
n(t_i) \sim \frac{\sqrt{1-a_0^2}\alpha t_i}{R}.
\end{equation}
We note that the condition $a_0^2 < 1$ ensures that $n(t_i) > 0 \forall t_i > 0$. Importantly one notes that the number of windings increases linearly
with time, with an overall coefficient modulated by the presence of the warp factor.

\n
Alternatively one can choose to solve for $w_l$ rather than the winding number;
\begin{equation} \label{eq:omega_l_2}
\omega_l \sim \sqrt{1-a_0^2}
\end{equation}
indicating that the magnitude of the warping imposes a physical constraint on the length of the string in the extra dimensions.
We can also identify this quantity with the velocity of the string in the compact space.
Imagine that the "end point" of the string at the horizon moves with a fixed velocity $\frac{v}{c} \leq 1$ in the extra dimensions. 
Then at time $t_i$ there will be approximately,
\begin{equation} \label{eq:n3}
n(t_i) \sim \frac{v t_i}{R}
\end{equation}
windings within the  horizon (with $c=1$) and the number of windings within a fraction $\alpha$ of the horizon is therefore, 
\begin{equation} \label{eq:n4}
n(t_i) \sim \frac{v \alpha t_i}{R}.
\end{equation}
This is equal to the number of windings per loop, for loops formed at time $t_i$ in the scaling regime. 
With this identification we see that
\begin{equation} \label{eq:omega_l_v}
\omega_l \sim v \sim \sqrt{1-a_0^2}
\end{equation}
and note the physical conditions $0 < v < 1$ and $0 < \omega_l < 1$ are automatically satisfied by the condition $0 < a_0^2 < 1$.
Furthermore it is intuitively obvious that $\omega_l$, $v, n(t_i) \rightarrow 0$ as $a_0^2 \rightarrow 1$ because the $a_0^2 = 1$ 
solution of the Klebanov-Strassler model corresponds to Minkowski space in $(9+1)$ dimensions. \footnote{Although this is not obvious from the 
formula (\ref{eq:a_0^2vsR^2}), the rationale behind this statement is the following; if no fluxes exist to provide an effective potential with which to 
compactify the extra dimensions, there can be no back reaction on the ordinary four-dimensional Minkowski manifold. 
Hence $a_0^2=1$ and $R^2 \rightarrow \infty$ automatically, leading to a flat six-dimensional space in place of the metric (\ref{eq:metric})/(\ref{eq:metric2}). Similarly the other dimensions of the bulk CY space are no longer flux-compactified. 
In such a scenario the formula (\ref{eq:a_0^2vsR^2}) would not be valid as, by definition, it holds only for $a_0^2 < 1$.} 
In this case there are no internal fluxes, and thus no compact extra dimensions implying that no windings can exist.

\n
We can also understand why $\omega_l$, $v \rightarrow 1$ as $a_0^2 \rightarrow 0$ if we realise that windings are effectively 
correlations which can only form \emph{within} the horizon. 
The horizon in the infinite dimensions advances according to the expression,
\begin{equation} \label{eq:d^{infty}_H}
d^{\infty}_H(t) =  a_0 t
\end{equation}
whereas in our background the horizon distance in Calabi-Yau space ($S^3$) is simply,
\begin{equation} \label{eq:d^{CY}_H}
d^{CY}_H(t) = t.
\end{equation}
Although the expression $a_0^2=0$ is strictly unphysical (corresponding to an extremal horizon), 
the limit $a_0^2 \rightarrow 0$ corresponds to a situation in which the 
infinite dimensions are "closed off" so that the string exists only in the compactified space (hence $\omega_l = 1$). 
In a time interval $\delta t$, the value of $a_0^2$ limits the increase of the horizon distance in the infinite directions (but not in the compact space) via,
\begin{equation} \label{eq:delta_d^{infty}_H}
\delta d^{\infty}_H = a_0 \delta t
\end{equation}
which places a limit on how fast the correlations can form. 
Strictly speaking, the horizon in the infinite direction advances by a distance $a_0  \delta t$ when the horizon in the compact space advances by $\delta t$. 
The end point of the string, which must of course move with resultant velocity $v_{res}=1$, must therefore cover a total 
distance \emph{in the compact space} $\delta d$ given by,
\begin{equation} \label{eq:delta_d}
\delta d= \sqrt{1 - a_0^2} \delta t
\end{equation}
which limits the effective velocity of the string in the extra dimensions to, $v \sim \frac{\delta d}{\delta t} = \sqrt{1-a_0^2}$. 
The parameter $\omega_l$ is then given simply by the ratio of the string velocity in the compact space 
to the speed of light, $\omega_l \sim v \sim \sqrt{1-a_0^2}$. 
Happily we find that the details of the compactification scheme determine $v$, $\omega_l$ and $n(t_i)$ uniquely 
\footnote{Note that in the preceding section it could be argued that, accounting for the effects of warping, the original formula for $n(t_i)$ 
in un-warped space (\ref{eq:n1}) (which was taken directly from \cite{Avgoustidis:2005vm}) 
should be modified to give $n(t_i) \sim \frac{\omega_l a_0 (\alpha t_i)}{R}$. However using this in conjunction with (\ref{eq:omega_l_1}) 
gives $n=\omega_l=0$ as the only possible solution. As there are good physical grounds (outlined above) for believing that the 
identification $\omega_l \sim v \sim \sqrt{1-a_0^2}$ (\ref{eq:omega_l_v}) is valid we therefore choose to leave the formula (\ref{eq:n1}) unchanged, 
even in the presence of warped space.}.
%%%%%%%%%%%%%%%%%%%%%%%%%%%%%%%%%%%%%%%%%%%%%%%%%%%%%%%%%%%%%%%%%%%%%%%%%%%%%%%%%%%%%%%%%%%%%%%%%%%%%%%%%%%%%%%%%%%%%%%%%%%%%%%%%%%%%%%%%%%%%%%%
\section{Comparison with the results of Blanco-Pillado and Iglesias}
We wish to consider a string loop which has windings over the full $S^3$. 
A previous study was initiated by \cite{BlancoPillado:2005dx} in which Euler variables were used to
describe the $S^3$ as an $SU(2)$ group manifold. This foliates the three
sphere into a one-parameter set of flat tori. For fixed angle $\theta_0$, the metric then reduces to that of a flat two-torus. The strings therefore 
only wrap a two-dimensional sub-manifold of the full three-sphere. We wish to generalize this result to consider windings over the full $S^3$ using
the Hopf map. We use the following ansatz to describe a string loop, with general, non-specific windings around the $S^3$
\begin{equation} \label{eq:ansatz}
X^M(\sigma,t) = \left(t, r(t)\sin(\sigma), r(t)\cos(\sigma), z_0, 0,0,0,0, \psi(\sigma,t), \theta(\sigma,t), \phi(\sigma,t)\right)
\end{equation}
where we have chosen our gauge so as to identify the world-sheet coordinate $\tau$ with the time coordinate in the Lorentz frame of the loop, 
i.e. $\tau \rightarrow t$. In keeping with the physical scenario we are considering, we specify the ansatz completely so that $\psi(\sigma,t), \ \theta(\sigma,t)$ and 
$\phi(\sigma,t)$ describe \emph{geodesic} windings. 
As previously stated, in Hopf coordinates the Killing vectors of the three-sphere take a
particularly simple form and are parallel to the unit vectors $(1,0,0)$, $(0,1,0)$ and $(0,0,1)$, so that geodesics windings are described by;
\begin{eqnarray} \label{eq:psi,theta,phi}
\psi(\sigma,t) &=& 2n_{\psi}\sigma + \psi(t) \nonumber\\
\theta(\sigma,t) &=& n_{\theta}\sigma + \theta(t) \\
\phi(\sigma,t) &=& n_{\phi}\sigma + \phi(t) \nonumber
\end{eqnarray}
where $n_{\psi},n_{\theta},n_{\phi} \in \mathbb{Z}$ represent the number of physical windings present in each angular direction. 
Note that the factor of two in front of the $n_{\psi}$ term is simply an artifact of the coordinate system, resulting from the fact 
that the principle range of the $\psi$-coordinate is twice that of $\theta$ and $\phi$ \footnote{The principle ranges of the angular coordinates 
are $0 \leq \psi < 2\pi$, $0 \leq \theta < \pi$, $0 \leq \phi < \pi$.}. 
Plugging (\ref{eq:ansatz}) and (\ref{eq:psi,theta,phi}) into the standard Nambu-Goto term of the $D$-string action 
(and ignoring the topological Chern-Simons term and other flux-dependent contributions) we have;
\begin{equation} \label{eq:action}
S = -T_1 \int d\sigma dt \sqrt{a_0^2(1 - \dot{r}^2)(a_0^2r^2 + R^2s'^2) - a_0^2r^2R^2\dot{s}^2}
\end{equation}
where $s = \psi + \theta + \phi$ and $T_1$ is the tension of the $D$-string. Rescaling the tension to absorb the string coupling allows us to also consider the
$F$-string. In the warped deformed conifold, the tension spectrum of a general $(p,q)$ string (in the large $p$ or large $q$ limit) is given by the well-known
formula \cite{Schwarz:1995dk,Firouzjahi:2006vp,Thomas:2006ud,Firouzjahi:2006xa}
\begin{equation} \label{eq:tension}
T_{(p,q)} = \frac{1}{2\pi \alpha'}\sqrt{\left(\frac{q}{g_s}\right) + \sin^2\left(\frac{p\pi}{M}\right)}
\end{equation}
which we assume to be valid for $p, q >>1$ and $p<<M$ in order for back-reaction effects to be neglected.
%\textcolor{red}{Note: I think that we originally missed out a factor of $a_0^2$ \emph{inside} 
%the square root of the Lagrangian. Looking back at my original calculations I think that what we had down as $(1 - \dot{r}^2)$ 
%should really have been $a_0^2(1 - \dot{r}^2)$. This does need to be checked more thoroughly though as it becomes important later on when 
%considering the "compatibility" of the results from section 1 (above) with Iglesias' arguments regarding the energy minimisation condition. 
%It is \emph{also} necessary to check the factor of $a_0^2$ in front of the $-r^2R^2\dot{s}^2$ term (though I also think this is correct) as it is also crucial later.}\\

\n
In fact we may simplify our result even further by an appropriate gauge choice with respect to the angular coordinates. 
Since we know that  geodesics of the compact space correspond to great circles on the $S^3$, we may set 
either of the winding numbers $n_{\theta}$ or $n_{\phi}$ to zero without loss of generality. 
The resulting string action with geodesic windings on the $S^3$ (in Hopf coordinates) is therefore,
\begin{equation} \label{eq:action2}
S = -T_1 \int d\sigma dt \sqrt{a_0^2(1 - \dot{r}^2)(a_0^2r^2 + R^2(2n_{\psi}+n_{\phi})^2) - a_0^2r^2R^2(\dot{\psi}+\dot{\phi})^2}.
\end{equation}
The resulting constants of motion are then;
\begin{equation} \label{eq:constants}
E = \frac{\partial L}{\partial\dot{q}^I}\dot{q}^I - L  \hspace{1.5cm} l = \frac{\partial L}{\partial \dot{q}^I}q'^I
\end{equation}
where $q^I \in \left\{t,\sigma,\psi,\theta,\phi \right\}$. 
The first expression is simply the Hamiltonian, and the second corresponds to the total angular momentum of the string in the compact directions 
\footnote{Note that although the string "rotates" around the $S^3$, no centripetal force is acting upon it. The internal (compact) dimensions are 
parameterised in terms of the angular variables $\psi$, $\theta$ and $\phi$, and so the motion through the $S^3$ may be measured in $rad \ \times \ [t]^{-1}$. 
As the effective radius of "rotation" for any point along the string is simply the radius of the three-sphere, 
multiplication by $R$ converts this "angular velocity" into the "true velocity" of the string. However, even here we must be careful - as the string has no internal structure, the "velocity" of the string parallel to itself (in this case parallel to the geodesic windings) is not clearly defined. It is therefore possible (in principle) for $v(t) \sim \dot{s}(t)R > 1$ though this does \emph{not} violate causality due to the boost invariance of the string along its length.}
Again using our ansatz (\ref{eq:ansatz})-(\ref{eq:psi,theta,phi}) we see that these expressions become:
\begin{eqnarray} \label{eq:constants2}
E &=& 2\pi T_1\frac{a_0^2(a_0^2r^2 + R^2(2n_{\psi}+n_{\theta})^2)}{\sqrt{a_0^2(1 - \dot{r}^2)(a_0^2r^2 + R^2(2n_{\psi}+n_{\phi})^2) 
- a_0^2r^2R^2(\dot{\psi}+\dot{\phi})^2}} \nonumber\\
l &=& 2\pi T_1\frac{a_0^2r^2R^2(2n_{\psi}+n_{\theta})(\dot{\psi}+\dot{\theta})}{\sqrt{a_0^2(1 - \dot{r}^2)(a_0^2r^2 + R^2(2n_{\psi}+n_{\phi})^2) 
- a_0^2r^2R^2(\dot{\psi}+\dot{\phi})^2}}.
\end{eqnarray}
Now Iglesias and Blanco-Pillado \cite{BlancoPillado:2005dx} have shown that, for a loop which is stationary in $3+1$ dimensions (i.e $\dot{r}=0$), 
the energy of the string configuration is minimised precisely for, 
\begin{equation} \label{eq:energymin1}
r^2 = \frac{l}{2\pi T_1 a_0^2}
\end{equation}
and 
\begin{equation} \label{eq:energymin2}
\dot{s}^2 = (\dot{\psi}+\dot{\theta})^2 = \frac{a_0^2}{R^2}.
\end{equation}
These results are obtained by first re-writing $\dot{s}^2=(\dot{\psi}+\dot{\theta})^2$ in terms of $l$, $r$, $R$ and $s'^2=(2n_{\psi}+n_{\theta})^2$ 
and then substituting into the expression for $E$ (so that $E=E(r,l,R,s')$), finally minimising with respect to $r$. 
Technically this gives the first condition ($\ref{eq:energymin1}$), and the second condition $(\ref{eq:energymin2})$ is then 
obtained by further substitution into the expression for $l$.

\n
Now any dynamical model we construct for the formation of geodesic windings \emph{and} for the motion of the string after loop formation must be 
consistent with these general results. At first sight we see that our model suggests $v^2(t_i) = \dot{s}^2(t_i)R^2 \sim (1-a_0^2)$, 
which does not correspond (in general) to the energy minimisation condition $v^2 = \dot{s}^2R^2 = a_0^2$. 
In fact these two conditions only coincide for the specific value $a_0^2 = 1/2$, where the velocity in both compact and non-compact dimensions is $v \sim 1/\sqrt{2}$.
We would not expect such a string configuration to undergo time evolution under the influence of its own internal dynamics, although it may still "shrink" 
via the loss of mass-energy (and angular momentum) due to gravitational wave emission and this possibility is discussed in section 6. 
We therefore conclude that if the value of the warp factor is exactly $a_0 = 1/\sqrt{2}$, the energy of the string configuration 
will be automatically minimised from the moment of loop formation i.e. the initial radius of the string loop $r(t_i)$ and the 
initial angular momentum $l(t_i)$ will be related via $r^2=l/2\pi T_1 a_0^2$, and the velocity of the string in the compact space 
will be $v \sim \sqrt{1-a_0^2} = a_0 = 1/\sqrt{2}$.
%\footnote{That our results are consistent with this may be verified directly 
%by substituting $a_0^2 = (1-a_0^2) = 1/2$ along with the expressions for $n(t_i)$ (\ref{eq:n}), $v = \dot{s}R$ (\ref{eq:omega_l_v})and $r(t_i)$ 
%into (\ref{eq:constants2}). This gives $l\alpha' = 1$ and $E\alpha' = a_0^2$ which, using that fact that $E\alpha' = a_0^2r^2$ implies 
%This is however only a specific case of the more general results...} 

\n
However when $a_0^2 > 1/2$, the string velocity in the compact space is too small to provide enough angular momentum to "match" the 
radius of the loop i.e. the angular momentum required for a cycloop of radius $r$ to minimise the energy of its configuration. Alternatively one can
understand this as a string loop with fixed energy, changing configuration in order to minimise the surface to energy ratio.
Hence from the arguments in Iglesias, we expect that $a_0^2 > 1/2$ implies $l < 2\pi T_1 a_0^2r^2$. 
Similarly if $a_0^2 < 1/2$ this should imply that the initial angular momentum of the loop exceeds the optimum value for a loop of that size and 
we find the converse result, namely $l > 2\pi T_1 a_0^2r^2$. In a full string theory compactification, the warp factor is exponentially small and
therefore this would appear to be the dominant string channel. However we will consider a more phenomenological approach in this paper, and consider
a range of values for the warping.
Furthermore, as stated above, these results should hold true for any physically viable dynamical model. 
It is therefore worth testing the theory developed in Section 2 to ensure consistency in this matter. 
In short Iglesias' and Blanco-Pillado's arguments \cite{BlancoPillado:2005dx} regarding the energy minimisation condition (for all $r$, $l$) 
ought to be consistent with our own dynamical model of $l(t_i)$ and $r(t_i)$ outlined above.

\n
%Using the order of magnitude estimate
Because we are dealing with geodesic windings, we may always re-define our coordinate system so as to identify the variable $n$ 
from (\ref{eq:n1}) with the variable $s'$, both of which represent the total number of \emph{physical} windings in the compact space - hence $n \sim 2n_{\psi}+n_{\theta}$. 
We now plug in the expressions for $n(t_i)$ (\ref{eq:n1}), $v^2 = \dot{s}^2R^2 = a_0^2$ (\ref{eq:energymin2}) and $r(t_i)=\alpha t_i$ 
into (\ref{eq:constants2}) (with $\dot{r}(t_i)=0$) to find the resulting terms \footnote{Note that we may could have substituted $n(t_i) = a_0^2 (\alpha t_i)^2/R^2$ 
in place of the usual expression $(\ref{eq:n1})$, taking advantage of the fact that $v^2 \sim a_0^2 R^2$ in this case. This leads to the expression 
$E(t_i) = 2\pi T_1 \times 2a_0^2(\alpha t_i)$ which is equivalent to (\ref{eq:constants_dynamical}) for $a_0^2 = (1-a_0^2) = 1/2$. 
We therefore see that the total energy is split half and half between the rest mass of the loop in warped Minkowski space, and the kinetic 
energy due to motion in the $S^3$.}
\begin{eqnarray} \label{eq:constants_dynamical}
E(t_i) &=& 2\pi T_1 \frac{a_0 (\alpha t_i)}{\sqrt{1-a_0^2}} \nonumber \\
l(t_i) &=& 2\pi T_1 a_0^2(\alpha t_i)^2 .
\end{eqnarray}
Hence we see that the second part of the energy minimisation condition (\ref{eq:energymin2}) implies the first, and vice-versa, as expected for consistency. 
However, as noted above, in general we have $v^2(t_i) = \dot{s}^2(t_i)R = (1-a_0^2)$ from (\ref{eq:omega_l_v}) which is not equal to $a_0^2$ unless $a_0^2=(1-a_0^2)=1/2$ 
giving the \emph{constant} values
\begin{eqnarray} \label{eq:constants_special}
E &=& E(t_i) = 2\pi T_1 (\alpha t_i) \nonumber\\
l &=& l(t_i) = \pi T_1(\alpha t_i)^2.
\end{eqnarray}
Under such special circumstances we would not expect the string configuration to evolve in time due to its own internal dynamics, 
though the emission of gravitational waves due to accelerated motion of the string, and the resulting shrinkage of the loop radius $r$ 
must still be accounted for as mentioned previously.

\n
Considering the more general case ($a_0^2 \neq 1/2$) and substituting $n(t_i)$ (\ref{eq:n1}), $v^2(t_i) = \dot{s}^2(t_i)R^2 = (1-a_0^2)$ 
(\ref{eq:omega_l_v}) and $r(t_i)=\alpha t_i$ into (\ref{eq:constants2}) - keeping $\dot{r}(t_i)=0$ - gives us
\begin{eqnarray} \label{eq:constants_general}
E(t_i) &=& 2\pi T_1 (\alpha t_i) \nonumber\\
l(t_i) &=& 2\pi T_1 (1-a_0^2)(\alpha t_i)^2
\end{eqnarray}
which implies that
\begin{eqnarray}
l(t_i) & > & 2\pi T_1 a_0^2(\alpha t_i)^2 \hspace{1.5cm} a_0^2 < \frac{1}{2} \nonumber \\
l(t_i) & < & 2 \pi T_1 a_0^2(\alpha t_i)^2 \hspace{1.5cm} a_0^2 > \frac{1}{2}.
\end{eqnarray}
\noindent
We find that the total energy of a cycloop with radius $r(t_i)=\alpha t_i$ is independent of $a_0^2$. 
At first glance this seems nonsensical: the value of the warp factor determines the velocity in the compact space at the moment of loop 
formation via $v(t_i) \sim \dot{s}(t_i)R \sim \sqrt{1-a_0^2}$, which in turn determines the initial number of 
loops via $n(t_i)R \sim v(t_i)r(t_i) \sim \sqrt{1-a_0^2}(\alpha t_i)$. 
A cycloop moving with greater velocity in the compact space would therefore have a greater number of windings than a slower moving string with the same radius
in the non-compact directions. 
Consequently an increase in the kinetic energy of the loop would seem to go hand in hand with an increase in the total rest mass. 
However although this is clearly true in \emph{un-warped} space, we must remember that the presence of the warp factor \emph{also} reduces the four dimensional
energy density via the effective tension $\tilde{T}_1 = a_0^2 T_1$. Equation (\ref{eq:constants_general}) suggests that even
though a smaller warp factor implies a greater rate of winding formation and a greater kinetic energy for the windings in the compact space, 
the would-be increase in the total energy of the cycloop is completely off-set by the reduction in four dimensional energy density.

\n
The question then remains: what happens if the energy minimisation conditions are not automatically satisfied at the moment of loop formation $t_i$? 
This is of course equivalent to the question: What happens dynamically when either $a_0^2 < 1/2$ or $a_0^2 > 1/2$? 
Intuitively we would expect that if $l(t_i) > 2\pi T_1 a_0^2 r(t_i)^2$ ($a_0^2 < 1/2$), the radius of the loop would rapidly expand 
introducing a non-zero $\dot{r}(t)$ term for $t > t_i$
\footnote{Note that this inequality is strict. At $t=t_i$ we still have that $\dot{r}(t_i)=0$. If this were not the case then the 
energy minimisation condition (\ref{eq:energymin1}) would itself be different.}.  
Physically this corresponds to the conversion of kinetic energy from the motion of the string in the compact space, into rest-mass energy in four dimensions. 
The conservation of angular momentum also suggests that any fraction of $l(t_i)$ 'lost' in this process (whatever this proportion may be) 
is carried away by the gravitational radiation produced by the the expanding loop, though as a first approximation we may neglect this.
%This in turn corresponds to a non-constant $\dot{r}$ term, as an accelerating loop is needed in order to generate gravitational waves. 
Hence we must allow for the most general case by including the explicitly time-dependent term $\dot{r}=\dot{r}(t)$ in the expressions for $E$ and $l$ 
(\ref{eq:constants2}), whose derivative $\ddot{r}(t)$ we expect to be initially positive for an expanding loop (i.e. $\ddot{r}(t_i)>0$).
Needless to say, we must also re-introduce a time-dependent velocity term $v(t) = \dot{s}(t)R$ for $t>t_i$, 
whose derivative $\dot{v}(t)$ we expect initially to be negative in this case ($\dot{v}(t_i) = \ddot{s}(t_i)R < 0$).

\n
Similarly if $l(t_i) < 2\pi T_1 a_0^2 r(t_i)^2$ ($a_0^2 > 1/2$) we expect the opposite process to occur - with rest-mass energy of the loop being 
converted into kinetic energy in the extra dimensions. For $t>t_i$ we again introduce the extra dynamical terms $\dot{r}(t)$, 
whose derivative $\ddot{r}(t)$ we now expect to be initially negative ($\ddot{r}(t_i)<0$) and $v(t) = \dot{s}(t)R$ whose derivative $\dot{v}(t)$ 
we now expect to be initially positive ($\dot{v}(t_i) = \ddot{s}(t_i)R > 0$). 
Again we face the possibility that a significant proportion of the initial angular momentum of the loop at the moment 
of formation will eventually be lost through the emission of gravitational radiation during dynamical evolution. 
%Technically therefore it is necessary, if the loop is eventually to reach a stable configuration, that the converstion of rest-mass energy into kinetic energy be such that the resulting gain in angular momentum outstrips the corresponding loss caused by the expansion of the loop in four dimensions. 
As a first approximation however, we will assume the loss of angular momentum via gravitational wave emission to be negligible, taking $l \approx l(t_i) \forall t>t_i$.
%\footnote{Footnote regarding the validity of this? Can we estimate the loss of energy/angular momentum via grav wave emission if we do \emph{not} assume that l remains constant? Is there any way to solve the equations explicitly if we take $E=E(t)$ and $l=l(t)$ explicitly? Is it even possible to generate grav waves which carry away some of the ang. mom. in the compact space via the movement of the string in the Minkowski directions? I think it is doubtful that it is, and if this is so, that argument by itself would be enough to justify the assumption $l=l(t_i) \forall t>t_i$}. 
We will also assume that the total energy lost via gravitational wave emission during the dynamical evolution of the loop is negligible 
i.e. that $E \approx E(t_i) \ \forall t>t_i$. What drives the evolution in this case is not changes in the energy of the system to "match" the conditions, 
but changes in the conditions to match the given energy: that is, the mutual and inter-dependent 
evolution of $r(t)$ and $\dot{s}(t)$ toward a loop configuration which meets the energy minimisation criteria (\ref{eq:energymin1})-(\ref{eq:energymin2}).

\n
The approach outlined above has the added advantage that in both cases we may assume that the number of windings remains fixed. 
As we shall see in the next section, the stability of the extra-dimensional windings places a lower bound on the value of $l$. 
If $l$ remains constant, all that is required to ensure stability of the windings throughout dynamical evolution towards the minimum energy state 
is that the string have sufficient angular momentum to stabilise its windings \emph{at the moment of loop creation}. 
In the next section we will demonstrate (following the analysis in \cite{BlancoPillado:2005dx}) that 
the stabilisation of windings places a bound on $l$, with more windings requiring a larger angular momentum to remain stable. 
Thus we see that, if the value of $l$ were to change significantly during the evolution process, the dynamics of the loop may be 
considerably more complicated with windings "falling off" the $S^3$ as the loop expands/contracts.    
%%%%%%%%%%%%%%%%%%%%%%%%%%%%%%%%%%%%%% 
\section{Loop dynamics after formation}
We now investigate the stability requirements for the extra-dimensional windings (as mentioned above) both generically and in light of our specific dynamical model. 
Happily we find that the rate of winding production in the model we have put forward ensures the stability of the ansatz (\ref{eq:ansatz})-(\ref{eq:psi,theta,phi}) 
at the moment of loop creation for all possible formation times $t_i$. We then leave this result on one side and consider the loop dynamics in each of the two 
different regimes ($l(t_i) > 2\pi T_1 a_0^2 r(t_i)^2$, $a_0^2 < 1/2$ and $l(t_i) < 2\pi T_1 a_0^2  r(t_i)^2$, $a_0^2 > 1/2$) discussed above on the assumption 
that the loops retain their windings during the evolution towards an energy-minimising state \footnote{Note that we will also find, in the next section, 
that the loop does not remain stable at the energy-mininimising configuration. However it is still true to say that it evolves from the initial radius \emph{towards} 
such a configuration. As we will show, the loops actually 'overshoots' its own energy-minimising configuration due to the non-zero velocity of the 
radial coordinate at that point, leading to an oscillating solution.}, which is equivalent to the assumption that $l=l(t_i)$ for all $t>t_i$.
%Finally we look for contradictions between the results derived using this assumption and the original constraints on $l$ from the condition of winding stability.\\

\n
By introducing a small perturbation in one of the bulk-space directions perpendicular to the $S^3$ it is possible to show, 
from the resulting expansion for the ten-dimensional action, that the string configuration (\ref{eq:ansatz})-(\ref{eq:psi,theta,phi}) 
is stable \cite{BlancoPillado:2005dx} (this corresponds to moving the entire string "up" from the very bottom of the throat by a small amount). 
The ansatz (\ref{eq:psi,theta,phi}) (with $n_{\theta}=0$) also implicitly assumes that the motion of the string in the compact space is parallel to the 
direction of the windings, so it is not physically meaningful to 'perturb' the string in either the $\psi$ or $\phi$-directions. 
We may however investigate the effect of perturbing the string in the $\theta$-direction in order to determine the stability of the 
winding configuration. Turning on a small perturbation $\delta \theta$ then results in the following perturbation of the Lagrangian \cite{BlancoPillado:2005dx}
\begin{equation} \label{eq:actionperturbation}
\delta L = \left(l/{2\pi T_1}a_0^2 + R^2(2n_{\psi}+n_{\theta})^2\right)\delta \dot{\theta}^2 - \frac{a_0^2}{R^2}\left(l/{2\pi T_1}a_0^2 - R^2(2n_{\psi}+n_{\theta})^2\right)\delta \theta^2
\end{equation}
which results in the stability condition
\begin{equation} \label{eq:stabilitycondn}
l >2\pi T_1 a_0^2R^2(2n_{\psi}+n_{\theta})^2
\end{equation}
In other words, the total angular momentum must satisfy this bound (note also the \emph{strictness} of the inequality) 
in order for the number of windings $(2n_{\psi}+n_{\theta})$ to remain stable. 
Again identifying $n \sim 2n_{\psi}+n_{\theta}$ and substituting for $n(t_i)$ (\ref{eq:n2}) and $l(t_i)$ (\ref{eq:constants_dynamical}), 
we may investigate the stability of the windings in our dynamical model \emph{at the moment of loop formation}. The resulting condition simply reduces to, 
\begin{equation} \label{eq:a_0^2<1}
a_0^2 < 1
\end{equation}
which is automatically satisfied by the definition of $a_0^2$ in the Klebanov-Strassler geometry. 
Thus we see that the stability condition is satisfied and that \emph{all} windings are stable at the time of loop 
formation for \emph{all} $t_i$ and for all physical values of $a_0^2$, $R^2$ and $\alpha$.

\n
We now introduce a non-zero time dependent term $\dot{r}(t)$ for $t>t_i$ in (\ref{eq:constants2}) which initially satisfies,
\begin{eqnarray} \label{eq:r_dot}
\ddot{r}(t_i) &>& 0 \hspace{1.5cm} a_0^2 < \frac{1}{2} \nonumber\\
\ddot{r}(t_i) &<& 0 \hspace{1.5cm} a_0^2 > \frac{1}{2}
\end{eqnarray}
and a time-dependent velocity term $v(t)=\dot{s}(t)R$ which initially satisfies,
\begin{eqnarray} \label{eq:v_dot}
\ddot{s}(t_i)R &=& \dot{v}(t_i) < 0 \hspace{1.5cm} a_0^2 < \frac{1}{2} \nonumber\\
\ddot{s}(t_i)R &=& \dot{v}(t_i) > 0 \hspace{1.5cm} a_0^2 > \frac{1}{2}
\end{eqnarray}
and which also satisfy the following boundary conditions, in the limit $t \rightarrow t_i$, 
\begin{eqnarray} \label{eq:r_and_v}
\dot{s}(t_i)R = v(t_i) \sim \sqrt{1-a_0^2}, \hspace{0.7cm} \dot{r}(t_i) = 0.
\end{eqnarray}
From now on then we may assume that the number of windings remains constant from the moment of loop formation, 
$n=n(t_i) \ \forall t>t_i$, and attempt to determine the corresponding dynamical evolution of the loop in the Minkowski directions. 
Using $l=l(t_i)$ and $E=E(t_i) \ \forall t>t_i$ we then have
\begin{eqnarray} \label{eq:l}
l = l(t_i) &=& 2\pi T_1 (1-a_0^2) (\alpha t_i)^2  \nonumber\\
&=& \frac{2\pi T_1 a_0^2 \sqrt{1-a_0^2} (\alpha t_i) r^2(t) \dot{s}(t) R}{\sqrt{a_0^2(1-\dot{r}^2(t))(a_0^2 r^2(t) + (1-a_0^2)(\alpha t_i)^2) - a_0^2 R^2 r^2(t) \dot{s}^2(t)}}
\end{eqnarray}
and
\begin{eqnarray} \label{eq:E}
E = E(t_i) &=& 2\pi T_1 (\alpha t_i) \nonumber\\
&=& \frac{2\pi T_1 a_0^2(a_0^2 r^2(t) + (1-a_0^2)(\alpha t_i)^2)}{\sqrt{a_0^2(1-\dot{r}^2(t))(a_0^2 r^2(t) + (1-a_0^2)(\alpha t_i)^2) - a_0^2 R^2 r^2(t) \dot{s}^2(t)}}
\end{eqnarray}
Re-arranging ($\ref{eq:l}$) then gives,
\begin{equation} \label{eq:s_dot_from_l}
a_0^2r^2\dot{s}^2R^2 = a_0^2(1-a_0^2)(1-\dot{r}^2) (\alpha t_i)^2
\end{equation}
and re-arranging ($\ref{eq:E}$) gives, 
\begin{equation} \label{eq:s_dot_from_E}
a_0^2 r^2 \dot{s}^2 R^2 = \frac{a_0^2(a_0^2 r^2(t) + (1-a_0^2)(\alpha t_i)^2)[(1-\dot{r}^2)(\alpha t_i)^2 - a_0^2(a_0^2 r^2(t) + (1-a_0^2)(\alpha t_i)^2)]}{(\alpha t_i)^2}
\end{equation}
so that equating the two expressions above yields the following non-linear first order differential equation in $r(t)$ 
\footnote{Alternatively of course we may substitute either of the expressions (\ref{eq:s_dot_from_E}) or (\ref{eq:s_dot_from_l}) into the 
original string action (\ref{eq:action2}) and then determine the Euler-Lagrange equations. The resulting equations must 
necessarily have the same solution as (\ref{eq:ODE_in_r}) but the method adopted here which utilises the string constants of motion is far simpler.}
\begin{equation} \label{eq:ODE_in_r}
\dot{r}^2 + \frac{a_0^4}{(\alpha t_i)^2} r^2 + (-1 + 2a_0^2(1-a_0^2)) + (1-a_0^2)^2 (\alpha t_i)^2 \frac{1}{r^2} = 0.
\end{equation}
We note that the constant terms involving $a_0^2$ and the terms involving powers of $r$ form a perfect square, so that this equation may be re-written as,
\begin{equation} \label{eq:perfect_sqr}
\dot{r}^2 -1 + \left(\frac{a_0^2}{(\alpha t_i)}r + \frac{(1-a_0^2)(\alpha t_i)}{r}\right)^2 = 0
\end{equation}
It is then explicitly clear that there exist two critical values of $r$ at which $\dot{r}=0$ i.e. at which the expansion of the 
loop (at least momentarily) comes to a halt. These are,
\begin{eqnarray} \label{eq:r_c1}
r_{c1} &=& (\alpha t_i)
\end{eqnarray}
and 
\begin{eqnarray} \label{eq:r_c2}
r_{c2} &=& \frac{(1-a_0^2)}{a_0^2} (\alpha t_i)
\end{eqnarray}
Although it is not possible to show this directly without the explicit form of the solution $r(t)$, we know that the first of these values 
must correspond to the boundary condition $\dot{r}(t_i)=0$ which we imposed when calculating $E(t_i)$ and $l(t_i)$, 
as well as when determining the energy-minimisation conditions (\ref{eq:energymin1})-(\ref{eq:energymin2}). 
However the second of these values is intriguing as it does \emph{not} correspond to the minimum energy condition. 
We can tell immediately therefore that the dynamical evolution of the loop will not lead to a steady energy-minimising state, 
and we may may instead expect a solution which oscillates between the two values $r_{c1} = (\alpha t_i)$ (the initial radius of the loop) 
and $r_{c2} = \frac{(1-a_0^2)}{a_0^2} (\alpha t_i)$.

\n
To see if such a solution is consistent with the physical arguments above i.e that a value of $a_0^2 < 1/2$ leads to an \emph{initially} expanding loop, 
$a_0^2 > 1/2$ leads to an \emph{initially} contracting loop and that $a_0^2= 1/2$ leads to a static loop of radius $r(t)=(\alpha t_i) \forall t>t_i$, 
we must now ask two questions. Firstly, which is greater, the initial radius or the second critical value? 
This will determine whether or not the loop initially expands or contracts. 
And secondly, which is greater, the second critical value or the radius corresponding to the energy-minimisation condition? 
This will determine whether the behaviour of the loop is in accordance with our assumptions.

\n
Whether $r_{c1} = r(t_i) <(>) r_{c2}$ therefore depends on whether $(1-a_0^2)/a_0^2 >(<) 1$ or equivalently whether $(1-a_0^2) >(<) a_0^2$. 
This in turn depends on whether $a_0^2 <(>) 1/2$ with $a_0^2 < 1/2$ implying that the loop must \emph{expand} from its initial 
value $r(t_i) = (\alpha t_i)$ to $r  = \frac{(1-a_0^2)}{a_0^2} (\alpha t_i) > (\alpha t_i)$ 
and $a_0^2 > 1/2$ implying that the loop must \emph{contract} from its initial value $r(t_i) = (\alpha t_i)$ to $r  = \frac{(1-a_0^2)}{a_0^2} (\alpha t_i) < (\alpha t_i)$.

\n
The second question may be answered as follows: By equating our expression for $l(t_i)$ (\ref{eq:constants_general}) 
with the second part of the energy-minimisation conditions (\ref{eq:energymin2}), we see that the critical radius 
corresponding to the fulfillment of this condition $r_{min}$ may be written in terms of $a_0^2$ and the initial radius $r(t_i)=(\alpha t_i)$ such that,
\begin{equation} \label{eq:r_min}
r_{min} = \frac{\sqrt{1-a_0^2}}{a_0}(\alpha t_i)
\end{equation}
Whether $r_{c2} <(>) r_{min}$ then depends on whether $(1-a_0^2)/a_0^2 <(>) \frac{\sqrt{1-a_0^2}}{a_0}$ 
which is equivalent to the condition $(1-a_0^2)/a_0^2 <(>) 1$ and reduces to $a_0^2 >(<) 1/2$.

\n
We therefore see that $a_0^2 < 1/2$ implies that the loop \emph{expands} from its initial value $r(t_i)=(\alpha t_i)$ \emph{towards} 
the radius corresponding to the energy-minimising configuration $r_{min} = \frac{\sqrt{1-a_0^2}}{a_0} (\alpha t_i) > r(t_i)$ 
but overshoots it and continues expanding to the second critical value $r_{c2} = \frac{(1-a_0^2)}{a_0^2}(\alpha t_i) > r_{min}$. 
Similarly $a_0^2 > 1/2$ implies that the loop \emph{contracts} from its initial value $r(t_i)=(\alpha t_i)$ \emph{towards} 
the radius corresponding to the energy-minimising configuration $r_{min} = \frac{\sqrt{1-a_0^2}}{a_0} (\alpha t_i) < r(t_i)$ 
but overshoots it and continues contracting to the second critical value $r_{c2} = \frac{(1-a_0^2)}{a_0^2}(\alpha t_i) < r_{min}$. 
In both cases, of course, we expect the loop to oscillate back and forth between $r_{c1}=r(t_i)$ and $r_{c2}$ 
but our findings are consistent with the assumptions about the initial behaviour of the loop expressed in the boundary conditions (\ref{eq:r_dot})-(\ref{eq:r_and_v}). 
We also note that for $a_0^2 = 1/2$, $r(t_i) = r_{c2} = r_{min} = (\alpha t_i)$ and the loop remains static in four dimensions.

\n
Having determined the general behaviour of our solution in each of the three cases $a_0^2 <=> 1/2$, we now return to equation (\ref{eq:ODE_in_r}) 
in order to find the explicit form of $r(t)$. By making the substitution $y=r^2$ (\ref{eq:ODE_in_r}) may be re-arranged to give,
\begin{eqnarray} \label{eq:integral}
dt = \pm \frac{1}{2} \frac{dy}{\sqrt{-ay^2 + by - c}}
\end{eqnarray} 
where we have defined the coefficients $a,b$ and $c$ to be explicitly positive for $a_0^2$ in the range $0<a_0^2<1$ so that,
\begin{eqnarray} \label{eq:abc}
a &=& \frac{a_0^4}{(\alpha t_i)^2} \nonumber\\
b &=& 1 - 2a_0^2(1-a_0^2) \nonumber\\
c &=& (1-a_0^2)^2 (\alpha t_i)^2
\end{eqnarray}
We also note for future reference that the discriminant of the quadratic $\Delta = b^2 - 4ac$ (given below) 
is positive for all $0<a_0^2<1$ except $a_0^2 = 1/2$ for which $\Delta = 0$.
%An integral of the form (\ref{eq:integral}) has several possible solutions depending on the sign of the coefficient $a$ and of the discriminant $\Delta = b^2 - 4ac$. In our case we see that $a = -a_0^4/(\alpha t_i)^2$ is always negative and that the discriminant $\Delta$ is given by,
\begin{equation} \label{eq:discriminant}
\Delta = b^2 - 4ac = 1 - 4a_0^2(1-a_0^2) = (1-2a_0^2)^2
\end{equation} 
Integrating (\ref{eq:integral}) then yields the following explicit expression for $t$ in terms of $y$, 
where $K$ is simply the usual integration constant and $A$, $B$ are the \emph{real} roots of the equation $-ay^2+by-c=0$
\footnote{This solution was obtained by performing a Euelerian transformation of the third kind. A brief sketch of this method is given in Appendix I.}
\begin{eqnarray} \label{eq:t}
t &=& \mp \frac{1}{\sqrt{a}} \tan^{-1}\left(\pm \sqrt{\frac{B-y}{y-A}}\right) \mp K
\end{eqnarray} 
which may be re-arranged to give $y=y(t)$,
\begin{equation} \label{eq:y_of_t}
y(t) = \frac{B + A \tan^2\left(\sqrt{a}(t+K)\right)}{1 + \tan^2\left(\sqrt{a}(t+K)\right)}
\end{equation}
Now the roots of the quadratic $-ay^2+by-c=0$ are given by,
\begin{eqnarray} \label{eq:roots}
y &=& (\alpha t_i)^2 \nonumber\\
y &=& \frac{(1-a_0^2)^2}{a_0^4} (\alpha t_i)^2
\end{eqnarray}  
but choosing $A=(\alpha t_i)^2$ and $B = \frac{(1-a_0^2)^2}{a_0^4} (\alpha t_i)^2$ does not allow us to fix $K$ in order to satisfy the boundary conditions. 
We must therefore choose $B=(\alpha t_i)^2$ and $A = \frac{(1-a_0^2)^2}{a_0^4} (\alpha t_i)^2$ in equation (\ref{eq:y_of_t}) 
before setting $y(t_i) = (\alpha t_i)^2$ to determine,
\begin{equation}
K=-t_i.
\end{equation}
Our final solution for $r(t)$ is then given by taking $r=+\sqrt{y}$ giving;
\begin{equation} \label{eq:r_solution}
r(t) = \sqrt{\frac{(\alpha t_i)^2 + \frac{(1-a_0^2)^2}{a_0^4}(\alpha t_i)^2 \tan^2\left(\frac{a_0^2}{(\alpha t_i)}(t-t_i)\right)}{1 + \tan^2\left(\frac{a_0^2}{(\alpha t_i)}(t-t_i)\right)}}
\end{equation}
which may be re-written in terms of basic trigonometric functions.
\begin{eqnarray}\label{eq:r_solution_2}
r(t) &=&  (\alpha t_i) \sqrt{1 + \left(\frac{1-2a_0^2}{a_0^4}\right) \sin^2\left(\frac{a_0^2}{(\alpha t_i)}(t-t_i)\right)}
\end{eqnarray}
It is clear that for $a_0^2 \neq 1/2$, the solution (\ref{eq:r_solution})/(\ref{eq:r_solution_2}) oscillates between $r(t_i) = (\alpha t_i)$ 
at $t = n\pi \frac{(\alpha t_i)}{a_0^2}$ for $n \in \left\{0,1,2,...\right\}$ and $r_{c2} = \frac{(1-a_0^2)}{a_0^2}(\alpha t_i)$ 
at $t = (2m+1)\frac{\pi}{2} \frac{(\alpha t_i)}{a_0^2}$, $m \in \left\{0,1,2,...\right\}$. 
We also see that for $a_0^2 = 1/2$ we have $A=B$ ($\Delta = 0$) and $r(t) = (\alpha t_i) \ \forall t>t_i$ as expected. 
It is also tedious but straightforward to twice differentiate (\ref{eq:r_solution}) in order to verify the boundary conditions (\ref{eq:r_dot}). The relevant quantities are,
\begin{eqnarray} \label{eq:r_dot_r_ddot}
\dot{r}(t)&=& \frac{(1-2a_0^2)}{a_0^2} \sin(x) \cos(x) \left\{1 + \left(\frac{1-2a_0^2}{a_0^4}\right) \sin^2(x)\right\}^{-\frac{1}{2}} \nonumber\\
\ddot{r}(t) &=& \frac{(1-2a_0^2)}{(\alpha t_i)} \left\{1 + \left(\frac{1-2a_0^2}{a_0^4}\right) \sin^2(x)\right\}^{-\frac{1}{2}} \\
&\times& \left[\cos^2(x)-\sin^2(x) - \left(\frac{1-2a_0^2}{a_0^4}\right) \sin^2(x)\cos^2(x)\left\{1 + \left(\frac{1-2a_0^2}{a_0^4}\right) \sin^2(x)\right\}^{-\frac{3}{2}} 
\right] \nonumber
\end{eqnarray}
where we have defined $x = \frac{a_0^2}{(\alpha t_i)}(t-t_i)$, from which it can clearly be seen that as $t \rightarrow t_i$ 
we have $\dot{r}(t_i)=0$ and $\ddot{r}(t_i)=\frac{(1-2a_0^2)}{(\alpha t_i)}$, and hence $\ddot{r}(t_i) > 0$ for $a_0^2 < 1/2$ and $\ddot{r}(t_i) < 0$ for $a_0^2 > 1/2$ in accordance with the boundary conditions (\ref{eq:r_dot}).
%Thus we see that all three physical possibilities, an initially expanding loop ($a_0^2 < 1/2$), 
%an initially contracting loop ($a_0^2 > 1/2$) and a constant static loop ($a_0^2 = 1/2$) are neatly encoded into a single mathematical formula.

\n
Figure 1  below  illustrates the qualitatively different behaviour of the solution (\ref{eq:r_solution_2}) 
for different values of $a_0^2$. The three curves show the three different types of dynamical evolution that a loop (formed with initial radius $r(t_i)=\alpha t_i$) may undergo in the $a_0^2 < 1/2$, $a_0^2 = 1/2$ and $a_0^2 > 1/2$ cases respectively.
%Figure 1
\begin{figure}[htp]
 \begin{center}
  \includegraphics[width=0.8\textwidth]{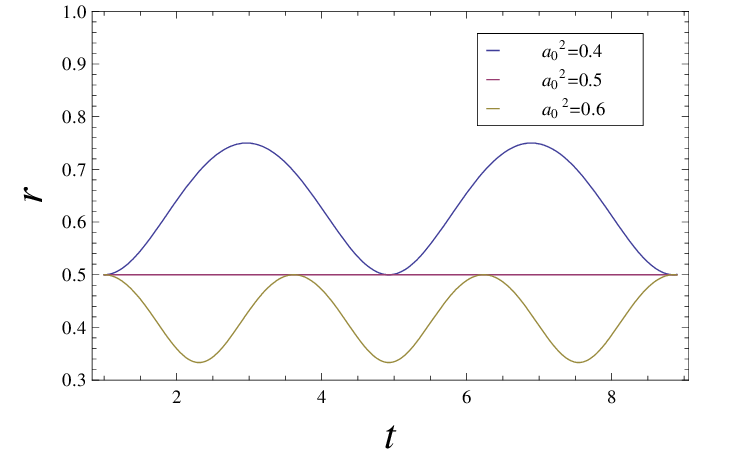}
  \caption{This figure illustrates the behaviour of the solution (\ref{eq:r_solution_2}) in the three qualitatively different regimes. 
For convenience we have chosen $t_i = 1$, $\alpha = 0.5$ for all three curves and set $a_0^2=0.4, 0.5$ and $0.6$ respectively.}
 \end{center}
\end{figure}
It is also worthwhile plotting the behaviour of $v(t) \sim \dot{s}(t)R$ for the same test values of $a_0^2$, $\alpha$ and $t_i$ which is shown in Figure 2. 
The formula $v(t)$ may be obtained by taking either (\ref{eq:s_dot_from_l}) or (\ref{eq:s_dot_from_E}) 
and substituting for $r(t)$ from (\ref{eq:r_solution_2}) and $\dot{r}(t)$ from (\ref{eq:r_dot_r_ddot}), 
though we will omit the explicit expression here for the sake of brevity. As expected, the behaviour of $v(t)$ (or rather $\dot{v}(t)$) clearly satisfies the 
boundary condition constraints (\ref{eq:v_dot}).
%Figure 2
\begin{figure}[htp]
 \begin{center}
  \includegraphics[width=0.8\textwidth]{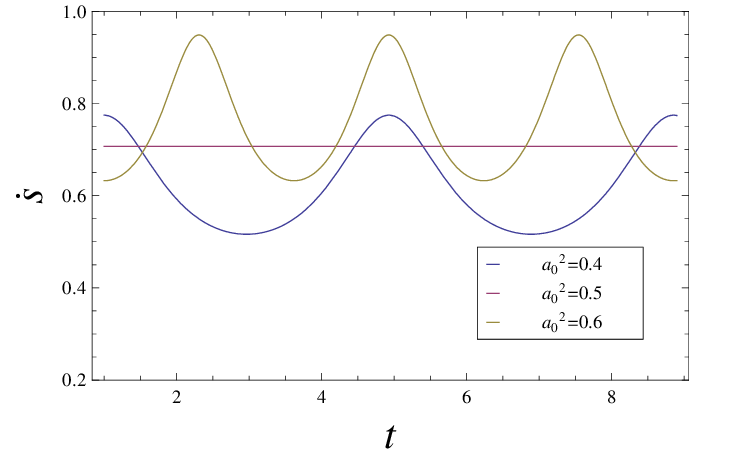}
  \caption{Figure illustrating the behaviour of $v(t) \sim \dot{s}(t)R$ in the three qualitatively different regimes. 
Again we have chosen $t_i = 1$, $\alpha = 0.5$ for all three curves and the values $a_0^2=0.4, 0.5$ and $0.6$ respectively.}
 \end{center}
\end{figure}
Even though, for any given model, the values of $a_0^2$ and $\alpha$ are constant; string loops are continuously formed throughout the history of the universe, 
meaning that $t_i$ will be a continuously changing parameter. In each of the oscillating regimes we would therefore expect to see a \emph{spectrum} 
of oscillation periods, with smaller loops formed at earlier epochs oscillating more rapidly between their initial and maximum radii than larger loops formed at late times. 
This is illustrated for the $a_0^2>1/2$ regime in Figure 3. The corresponding behaviour of $v(t) \sim \dot{s}(t)R$ is shown in Figure 4 using the same 
test values for the parameters.  
%Figure 3
\begin{figure}[htp]
 \begin{center}
  \includegraphics[width=0.8\textwidth]{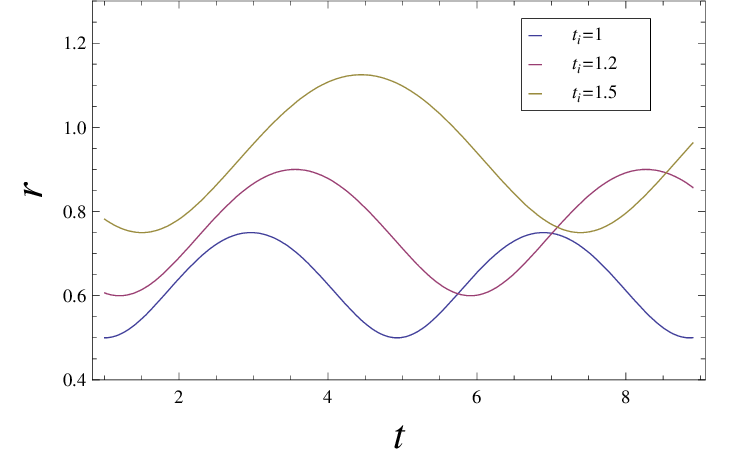}
  \caption{This figure illustrates the behaviour of $r(t)$ for loops formed at three different epochs in the $a_0^2>1/2$ regime. 
For the sake of convenience we have fixed $a_0^2 = 0.4$ and $\alpha = 0.5$ for all three curves and set $t_i=1$, $t_i=1.2$, and $t_i=1.5$ respectively.}
 \end{center}
\end{figure}
%Figure 4
\begin{figure}[htp]
 \begin{center}
  \includegraphics[width=0.8\textwidth]{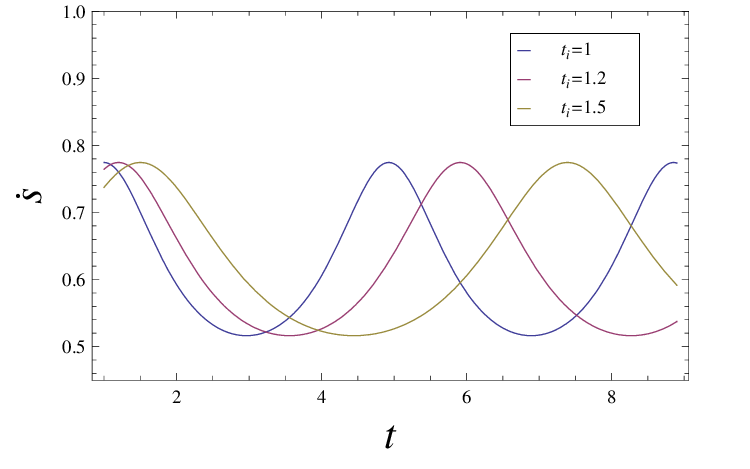}
  \caption{This figure illustrates the behaviour of $v(t) \sim \dot{s}(t)R$ for loops formed at three different epochs in the $a_0^2>1/2$ regime. 
As in Figure 3 we choose to set $a_0^2 = 0.4$ and $\alpha = 0.5$ for all three curves and consider $t_i=1$, $t_i=1.2$, and $t_i=1.5$.}
 \end{center}
\end{figure}
%*****************************************************************************************************************************************************************
\section{Discussion}
We have argued that a velocity correlations regime in the post-inflationary epoch leads naturally to the formation of string loops with geodesic 
windings in the compact space.  For strings at the tip of the conifold throat of the Klebanov-Strassler geometry 
we were able to show that the quantities which determine the dynamical evolution of the loop 
(i.e. the initial winding number $n(t_i)$, energy $E(t_i)$ and string velocity/angular momentum in the compact space, $\dot{s}(t_i)$/ $l(t_i)$) 
are \emph{uniquely} determined by the parameters $a_0^2$, $\alpha$ and $t_i$. 
Crucially these windings were found to have sufficient angular momentum in the compact directions to remain stable, 
even after the string chops off from the network to form a loop. 

\n
The interaction between the tension and the angular momentum in the compact space was found to play a significant role in the dynamical 
evolution of the string, including - perhaps surprisingly - the evolution in four dimensions.  
By assuming energy loss (and angular momentum) via gravitational radiation to be negligible, 
we determined equations of motion for the four-dimensional radius $r(t)$ and the string velocity $v(t) \sim \dot{s}(t)R$, 
which we believe to be valid over small time scales after the moment of loop formation, $t=t_i$. 

\n
We found that the qualitative behaviour of the string depends crucially on the square of the warp factor, $0<a_0^2<1$, with $a_0^2<1/2$ leading to an 
oscillatory solution characterised by an initial phase of expansion, whilst $a_0^2>1/2$ leads to oscillations with an initial contracting phase. 
In each case the string was seen to oscillate between it's initial radius $r(t_i)=(\alpha t_i)$ 
and a secondary critical value defined by $ r_{c2}=\frac{(1-a_0^2)}{a_0^2}(\alpha t_i)$, with the period of oscillation given by $T = \frac{(\alpha t_i)} {a_0^2}$. 

\n
As noted above, in the two oscillatory regimes it is the interaction of the angular momentum with the string tension which 
"drives" the dynamical evolution converting kinetic energy into rest mass during expansion (with the inverse process occurring in the contracting phase).
The string is seen to evolve \emph{towards} a static, minimum energy configuration where $l=2\pi a_0^2 T_1$ and $\dot{s}^2 = a_0^2/R^2$, 
but is unable to satisfy these two conditions simultaneously at a point where $\dot{r}(t)=0$. 
By contrast, in the $a_0^2=1/2$ scenario we find that the energy minimising conditions \emph{are} satisfied simultaneously at the moment of loop formation. 
In this case the tension exactly offsets the effect of angular momentum so that the string remains static at it's original radius $r(t_i)=(\alpha t_i)$ 
and is non-oscillatory.

\n
The meaning of the term "small" above is somewhat ambiguous, but it seems reasonable to assume that our solution will provide a valid approximation 
over at least one full oscillatory cycle of the loop; that is, over a time period $\Delta t \sim T = \frac{(\alpha t_i)}{a_0^2}$. 
For time periods $\Delta t >> T$ our initial analysis must be extended to include the effects of gravitational wave emission on $E(t)$ and $l(t)$ 
(or equivalently on $r(t)$ and $\dot{s}(t)$). 
We may expect the qualitative effect of energy and angular momentum loss to be relatively simple, as long as the string retains sufficient angular momentum 
for the extra-dimensional windings to remain stable. In this case it is likely that the loss of $E$ and $l$ due to gravitational wave emission 
will simply act to damp the oscillations of $r(t)$ and $\dot{s}(t)$. 
What is unclear however is whether or not the damping coefficient itself is likely to be time-dependent; 
For example, it is possible that smaller oscillations lead to greater rates of emission per unit length 
(as the rate of acceleration $\ddot{r}(t)$ is higher in this case) so that the damping itself increases with time (for an individual loop).

\n
In the case that the windings eventually become unstable however (as $l(t)$ drops below the threshold for ensuring their stability) 
the string dynamics are likely to become complicated, and it is not clear whether the process of winding contraction 
(i.e. of windings "falling off" the $S^3$) may even be accommodated within an analysis which uses an ansatz of the form (\ref{eq:ansatz}) 
to describe the string configuration. 
This is because the coordinates $r(t)sin\sigma$, $r(t)cos\sigma$, $\psi(\sigma,t)$, $\theta(\sigma,t)$ and $\phi(\sigma,t)$ 
are treated as \emph{independent} variables with respect to the determination of the equations of motion. 
We are therefore unable to take account of the continuously connected nature of the string when string sections "move" from  
one direction to another (e.g. in "falling off" the $S^3$ to form part of the four dimensional rest mass of the loop).

\n
The present analysis could of course still be improved by accounting for the effects of gravitational wave emission under the 
assumption that that loops retain their windings, which would indeed be valid up to the point where $l(t)$ drops below the critical value 
(which may also be calculated). Such an improved analysis could proceed as follows: 
one could compute the stress energy tensor for an oscillating loop and look for solutions with this as 
a source to the Einstein equations. 
This should allow us to estimate the rate of loss of $E$ and $l$ via gravitational wave emission and, as stated, 
we expect this to produce a damping term in the equations for $r(t)$ and $\dot{s}(t)$. 

\n
Calculating the emission spectrum for an oscillating loop would also be of immense practical interest. 
The gravitational wave signature of such a loop, whose self-oscillation is caused by the presence of angular momentum in the compact space, 
may differ significantly from that of a string whose oscillations, though superficially similar, 
do not result from self-interaction. 
Gravitational wave emission from loops oscillating with period $\omega \sim L^{-1}$ (where $L$ is the loop size) 
have been intensively studied in four dimensions \cite{Vaschpati+Vilenkin:1985, Burden:1985, Garfinkle+Vaschpati:1987, Allen+Shellard:1992}. 
However in such cases the loop is not undergoing \emph{genuine} phases of expansion and contraction, but rather experiencing "wiggles" of a size 
comparable to it's own length. 
Although Weinberg \cite{Weinberg:1972} has has shown that - in an FRW universe - the power of a weak, isolated, periodic source 
(to lowest order in $G$) may be given by a single formula \emph{regardless} of the exact nature of the source, 
it is not immediately clear that this should hold in extra-dimensional scenarios. 
Additionally such sources (i.e. loops) have no angular momentum to shed in the process of emission. 
In fact even if we were to study loops whose self-oscillation was due to their "rotational" motion in Minkowski space 
(c.f. \cite{Durrer:1989, Vilenkin+Shellard:2000}), the angular momentum which would be carried away by gravitational radiation 
would be very different from that lost via emission from oscillating cycloops 
\footnote{We note also that one possible further extension of the current analysis would be to consider a wound string with rotational 
motion in both the compact \emph{and} Minkowski directions.}. 
The detection of such a signature by the future generation gravitational wave detectors
could then provide indirect evidence of the presence of compact dimensions. We hope therefore to be able to provide such an analysis in the near future.

\n
Although the full string theory compactification favours exponentially small values of $a_0^2$, 
it is also interesting to note that the possibility of obtaining evidence from the gravitational wave signature of 
wound strings exists even in the case of static loops (i.e. the $a_0^2 = 1/2$ case). 
These loops, which form automatically with a configuration which meets the energy minimising conditions, 
look just like ordinary string loops from a four-dimensional perspective. 
However they contain an "unseen" angular momentum which is not directly manifested in their dynamical evolution in Minkowski space. 
As stated above, we expect that even "standard" four dimensional string loops may undergo periodic oscillations with $\omega \sim L^{-1}$ 
creating ripples in space-time and giving rise to gravitational waves. 
Such fluctuations \emph{along} the length of the string - though not in the total four dimensional length itself - 
would likely still occur in this case (though it is possible that the existence of the angular momentum term may lend a certain 'rigidity' 
to the circular string configuration, making it more resistant to deformation) 
but the string must also now shed it's angular momentum. 
The gravitational wave signature of even a non-self-oscillating loop in the extra-dimensional scenario is therefore also likely to differ 
significantly from the standard case of an un-wound string, 
though perhaps to a less significant degree than in the self-oscillating case. 
However the possibility of the indirect detection of compact dimensions from cosmic strings therefore remains, 
even in the absence of self-oscillating loops. We hope also to be able to offer an analysis of this interesting possibility in a future letter. 

\n
It remains for us to outline some of the possible limitations of the analysis we have attempted: 
We have assumed throughout the present work that the velocity correlations regime leads naturally to 
a) geodesic windings and b) movement of the string parallel to itself (i.e. along the geodesics). 
However these assumptions may be questioned. 
The rationale for adopting such an approach (which simplified the resulting analysis considerably) was simple - 
we expect velocity correlations to impart a constant angular momentum density to each point along the string. 
Therefore we expect each "point" (or infinitesimal string segment) to travel along a geodesic curve, both before and after loop formation. 
However it has not been proved that each point along the string traversing a separate geodesic, 
leads to a winding configuration that is itself geodesic. 
Likewise it does not necessarily follow that the resulting motion of the string as a whole is parallel to itself. 
In general, we may expect that the geometry of the internal space plays a role in determining the 
exact nature of the winding configuration and the resulting string motion. 
Moreover we have not included the contribution from the Ramond-Ramond (RR) sector, which in principle will couple to the string. This charge
term is ultimately crucial for distinguishing between cosmic strings and cosmic super-strings.

\n
Whilst, in principle, motion of the string perpendicular to its length may easily be accounted for 
(see end of Appendix II) the most significant problem in the analysis of string loops with non-geodesic windings 
arises from the resulting $\sigma$-dependence of the expressions for $E$ and $l$. 
We believe that the present analysis may therefore be improved by a more thorough investigation of the winding process itself, and though we expect 
our expressions for $n(t_i)$ (\ref{eq:n2}) and $v(t_i) \sim \dot{s}(t_i)R$ (\ref{eq:omega_l_v}) to remain valid (in the KS case), 
the appropriate string ansatz may well be more complicated. 
Although it seems unlikely that the qualitative behaviour of the string will differ significantly than 
described in the scenarios above, finding analogous (and \emph{quantitatively} different) results may be extremely difficult.
But ultimately exact \emph{quantitative} predictions will be needed for any comparison with future experimental data, and we believe such a project to be worthwhile.  
%%%%%%%%%%%%%%%%%%%%%%%%%%%%%%%%%%%%%%%%%%%%%%%%%%%%%%%%%%%%%%%%%%%%%%%%%
\begin{center}
{\bf Acknowledgments}
\end{center}
M.L is supported by an STFC studentship and the Queen Mary EPSTAR consortium.
This work was supported in part by NSERC of Canada. 
%*****************************************************************************************************************************************************************
\renewcommand{\theequation}{A-\arabic{equation}}
\setcounter{equation}{0}  % reset counter 
\section*{Appendix I: Eulerian substitution of the third kind}
As noted previously, the quadratic equation in the integral defined by (\ref{eq:integral}) and (\ref{eq:abc}) 
has a discriminant $\Delta = b^2 - 4ac$ which is everywhere non-negative for values of $a_0^2$ in the range $0 < a_0^2 < 1$. 
We may therefore evaluate the integral using a Eulerian transformation of the third kind, which takes advantage of the fact that there exist 
two \emph{real} roots in $y$. We proceed in the following way: 
Let $A$ and $B$ be the two real roots of the quadratic equation $-ay^2 + by - c = 0$. We then define a dummy variable $u$ such that,
\begin{equation} \label{eq:E1}
\sqrt{-ay^2 + by - c} = (y-A)u
\end{equation}
In general of course we have that,
\begin{equation} \label{eq:E2}
-ay^2 + by - c = a(y-A)(B-y)
\end{equation} 
so that combining (\ref{eq:E1}) and (\ref{eq:E2}) implies,
\begin{equation} \label{eq:E2a}
a(B-y)=(y-A) u^2 
\end{equation}
or equivalently,
\begin{equation} \label{eq:E3}
y = \frac{aB+Au^2}{a+u^2}
\end{equation}
and 
\begin{equation} \label{eq:E3a}
\frac{y}{\sqrt{a}} = \pm \sqrt{\frac{B-y}{y-A}}
\end{equation}
Using the second solution for $y$ as function of $u$ and differentiating the expression above then gives,
\begin{equation} \label{eq:E4}
dy = \frac{2du}{a + u^2}\left\{-\left(\frac{aB+Au^2}{a+u^2}\right) + A\right\}u
\end{equation}
Finally, substituting (\ref{eq:E1}), (\ref{eq:E3}) and (\ref{eq:E4}) into the right hand side of (\ref{eq:integral}) gives
\begin{eqnarray} \label{eq:E5}
\int dt &=& \pm \frac{1}{2} \int \frac{dy}{\sqrt{-ay^2 + by - c}} \nonumber\\
&=& \pm \frac{1}{2} \int \frac{dy}{(y-A)u} \nonumber\\
&=& \pm \frac{1}{2} \int \frac{\frac{2du}{a+u^2} \left\{-\left(\frac{aB+Au^2}{a+u^2}\right) + A\right\}u}{\left\{\left(\frac{aB+Au^2}{a+u^2}\right) - A\right\}u} \nonumber\\
&=& \mp \int \frac{du}{a+u^2}
\end{eqnarray}
Using the standard integral $\int \frac{dx}{\kappa^2 + x^2} = tan^{-1}\left(\frac{x}{\kappa}\right)$ and (\ref{eq:E3a}) then gives equation (\ref{eq:t}).
%*****************************************************************************************************************************************************************
\renewcommand{\theequation}{B-\arabic{equation}}
\setcounter{equation}{0}  % reset counter 
\section*{Appendix II: The Hopf fibration of the three-sphere}
As stated in the introduction, we chose to use the Hopf fibration of the $S^3$ when considering geodesic windings, as this allows the 
metric and Killing vectors to be written in a particularly simple form viz (\ref{eq:metric2})/(\ref{eq:psi,theta,phi}). 
Choosing windings which wrap only Killing directions in the $S^3$ then leads to \emph{manifest} $\sigma$ independence in the constants of motion (\ref{eq:constants2}). 
The ansatz corresponding specifically to \emph{geodesic} windings is the linear ansatz (\ref{eq:psi,theta,phi}), 
and the choice of coordinates naturally reflects the $SO(3)$ symmetry of the internal space. 

\n
We will now discuss the origin of the Hopf fibration from a geometric point of view.
However it will also be useful to discuss the origin of the canonical coordinate system from a similar perspective. 
By comparing the two (equivalent) descriptions, we hope to make clear the advantage of using the former coordinate system in the present work, 
but also to demonstrate explicitly the coordinate independence of our results.

\n
One very natural choice of coordinates on the $S^3$ is  the so-called 'canonical parameterisation'. 
In this case the group element of the $S^3$ manifold is given by,
\begin{equation}
g = e^{(i/2)[x \sigma_1 + y \sigma_2 +z\sigma_3 ] },
\end{equation}
where $\sigma_i$, $i \in \left\{1,2,3\right\}$ are the usual Pauli matrices. 
Writing the usual Cartesian coords $x, y $ and $z$ in terms of polars $r, \theta, \phi $ the line element then becomes,
\begin{equation}
ds^2 = -\frac{1}{2}  Tr ([dg g^{-1}]^2 ) = dr^2 + \sin^2r (d\theta^2 + \sin^2\theta d\phi^2)
\end{equation}
which is exactly that given in (\ref{eq:metric}) if we identify $\psi = r$ and multiply the whole metric by $R^2$.

\n
By contrast, and following Iglesias and Blanco-Pillado \cite{BlancoPillado:2005dx}, 
we chose to parameterise the $S^3$ using Eulerian variables. In this case we regard the $S^3$ as an $SU(2)$ group manifold with element $g$ given by,
\begin{equation}
 g(\psi, \theta, \phi) = e^{i(\psi/2)\sigma_1} e^{i(\theta/2 )\sigma_2} e^{i(\phi/2 )\sigma_3 }
\end{equation}
where $\sigma_i$ are the Pauli matrices as before. The group invariant metric (i.e. the metric on $S^3$ in these coordinates) can then be written in the following
form
\begin{equation}									
ds^2 =  -\frac{1}{2} Tr ({[dgg^{-1}]}^2) = \frac{1}{2} (d\psi + \cos\theta d\phi)^2 + \frac{1}{2}(d\theta)^2.
\end{equation}
It may then be seen that, in these coordinates, the sub-manifolds where $\theta = \theta_0 $ is constant correspond to flat two-tori with metrics given by,
\begin{equation}
ds^2 = \frac{1}{2}(d\psi + \cos\theta_0 d\phi)^2
\end{equation}
This shows that the Hopf fibration corresponds to describing the three-sphere as a one-parameter family of flat two-tori (to which it is topologically equivalent). 
Again we follow Iglesias \cite{BlancoPillado:2005dx} in choosing $\theta_0 = 0 $ in order to simplify the metric as much as possible. 
It is of course also necessary to re-scale the metric so that $1/2 \rightarrow R^2$ as we are dealing with a physical $S^3$ of radius $R$. 

\n
Let us now consider an arbitrary string embedding which is a function of both space and time, $X^{i}(\sigma,t)$ where $i \in \left\{0,1, \ldots 10\right\}$. 
The general - coordinate independent - expression for the string Lagrangian is then,
\begin{equation}\label{generalaction}
L = -T_1 a_0^2 \int d\sigma \sqrt{(1-\dot{r}^2)(r^2+a_0^{-2}R^2 W)- a_0^{-2}R^2 r^2 P + a_0^{-4} R^4 Q}
\end{equation}
where
\begin{eqnarray}
W &=&  {X}'^i(\sigma, t) g_{ij}(X(\sigma, t)){X}'^j(\sigma,t)\\ \nonumber
P &=& \dot{X}^i(\sigma,t) g_{ij}(X(\sigma,t))\dot{X}^j(\sigma,t)\\ \nonumber
Q&=&(\dot{X}^i(\sigma,t) g_{ij}(X(\sigma,t)){X}'^j(\sigma,t))^2 \\ \nonumber 
&-& (\dot{X}^i(\sigma,t) g_{ij}(X(\sigma,t))\dot{X}^j(\sigma,t))({X}'^i(\sigma,t) g_{ij}(X(\sigma,t)){X}'^j(\sigma,t)) 
\end{eqnarray}
where a dash indicates differentiation with respect to $\sigma$ and a dot represents differentiation with respect to $t$ as usual. 
In the static case the term $W(\sigma) = {X}'^i(\sigma) g_{ij}(X(\sigma)){X}'^j(\sigma)$ corresponds simply to the length of the 
wrapped string on the $S^3$. Demanding that $\frac{dW}{d\sigma} =0$ one simply obtains the geodesic equation,
\begin{equation}
X''^i(\sigma) +  \Gamma^i_{jk}X'^j(\sigma )X'^k(\sigma)  =0
\end{equation}
where $\Gamma^i_{jk}$ are the usual Christoffel symbols. Thus minimal (i.e. geodesic) windings on a general manifold (in our case this is just $S^3$) with metric $g_{ij}$ is enough to guarantee 
$\sigma$ independence of the winding terms.

\n
The same statement still holds true if we allow time-dependence into the string ansatz, $X^i(\sigma,t)$, as long as for each instant in time
the latter still satisfies the geodesic equation. Physically this means that the wrapped loop only ever evolves along geodesic curves. 
A simple example on the $S^2$ would be a great circle passing through the north and south poles and rotating about an axis through the poles.   
Thus we require our embedding function to satisfy,
\begin{equation}\label{geod}
X''^i(\sigma,t) +  \Gamma^i_{jk}X'^j(\sigma,t )X'^k(\sigma,t)  =0
\end{equation}
Now, with time-dependence we also have a non-zero kinetic energy term of the wrapped string inside the square root factor in our Lagrangian,
\begin{equation}
P(\sigma, t) =  \dot{X}^i(\sigma,t) g_{ij}(X(\sigma,t))\dot{X}^j(\sigma,t).
\end{equation}
Demanding now that $\frac{dP}{d\sigma} =0 $ we find,
\begin{equation}\label{preserv}
\dot{X}'^i(\sigma, t)  +  \Gamma^i_{jk}\dot{X}^j(\sigma, t )X'^k(\sigma, t)  =0
\end{equation}
Geometrically this equation states that the velocity vector, $\dot{X}^i$, is preserved under parallel transport along a geodesic curve that is 
wrapped by the string. One can equally interpret this equation as saying that under parallel transport of the tangent vector 
to the geodesic curve, $X'^i$, along a curve whose tangent vector is $\dot{X}^i$, the former is preserved. 
Mathematically this is simply the statement that $\nabla_{\dot{\gamma}} \gamma' \equiv \nabla_{\gamma'} \dot{\gamma} = 0$.

\n
The nice consequence of demanding the above is that that the winding term is not only $\sigma$ independent (if we take geodesic wrapping) 
but also time independent, i.e. $\frac{dW}{dt} =0$. Thus $\sigma$ independence of the kinetic function $P$ guarantees time independence of $W$.
This is an important statement.

\n
Let us now consider what the situation above implies for the case where the wrapping is over flat sub-manifolds as in
the case of Iglesias, where the $\Gamma^i_{jk}$ all vanish. 
The geodesic equation (\ref{geod}) and (\ref{preserv}) trivially imply $X''^i =0, \dot{X}'^i =0$, and therefore the solution of these equations is,
\begin{equation} \label{eq:appendix_linear_ansatz}
X^i(\sigma,t) = X^i_0 + n^i \sigma +  u^i(t).
\end{equation}
This provides the origin of the linear $\sigma$ anasatz of Iglesias. Now apart from the winding and kinetic terms mentioned above there are 
additionally the two $R^4$ terms appearing inside the square root in the Lagrangian in $Q(\sigma, t)$. 
The $\sigma$ independence of the second term follows from our previous results, because this term is simply $W P$. 
The $\sigma$ independence of the first term above follows after a simple calculation making use of  (\ref{geod}) and (\ref{preserv}). 
In the case of wrapping along a flat sub-manifold one sees that $Q$ vanishes identically. 
%\textcolor{red}{For the general case I haven't been able to prove this still happens, using (\ref{geod}) and (\ref{preserv}). 
%It is perhaps not so important if this term vanishes or not.}

\n
However what is important is that $\sigma$ independence of all the relevant terms in the Lagrangian is guaranteed simply by 
requiring the strings to wrap geodesics and that the velocity vector is preserved under parallel transport along this geodesic. 
Thus we see that $L = \int d\sigma \mathcal{L} = 2\pi \mathcal{L}$ as stated in the introduction.

\n
We now have the possibility of working in \emph{any} coordinate system, though the choice of explicit string wrapping ansatz is 
constrained by the requirement of solving (\ref{geod}), (\ref{preserv}) \emph{if} we require geodesic windings. 
In general explicit solutions to the geodesic equations are hard to come by but for very symmetric spaces - like spheres - they are known 
\footnote{For general wrappings of the string around the full $S^3$ however, the ansatz will certainly never be linear in $\sigma$ and nor will the $\sigma$ 
and time dependence factorize in an additive way as it did in the flat space case, no matter what choice of coordinates we make.}.

\n
However let us now compare the form of these explicit solutions, in canonical coordinates, to the simple linear ansatz in (\ref{eq:appendix_linear_ansatz}). 
In order to determine the appropriate ansatz for geodesic windings in canonical coordinates we must first calculate the Killing vectors in this coordinate system. 
We know that the Killing fields corresponding to the $SU(2)$ rotations of the $S^3$ generate isometries of the above metric - therefore, if we think of the $S^3$ as 
being embedded in $R^4$ where ${X,Y,Z,U}$ represent the four Cartesian coordinates,
\begin{equation}
X^2 +Y^2 +U^2 +Z^2 =1,
\end{equation}
the three independent isometries are generated by 
\begin{equation}
J_1 = J_{XU} + J_{YZ}, \qquad J_2 = J_{XZ} + J_{YU}, \qquad J_3 = J_{XY} + J_{ZU}
\end{equation}
where $J_{ZU} = Z\partial_U - U\partial_Z  $ is the  generator of rotations in the $Z-U$ plane etc.
The above $J_i $ clearly generate an $SU(2)$ algebra.
Then the three Killing vectors are,
\begin{equation}
k^i_1 = (-U,-Z,X,Y); \qquad  k^i_2 = (-Z,U,-Y,X) ; \qquad k^i_3 = (-Y,X,Z,-U)  
\end{equation}
which, in fact, define an orthonormal basis for the $SU(2)$ Lie algebra. 
However \emph{unlike} the Killing vector fields of the two-torus, those of the whole $S^3 $ are in general coordinate dependent.

\n
Since we work with coords $\psi,\theta,\phi $ rather than the embedding coords $X,Y,U,Z $
one can re-express the above Killing vectors in terms of $k^i_a (\psi, \theta, \phi ) $
with $ i=1 \ldots 3 $ of the $S^3 $. To obtain these just consider the left action of rigid group elements
\begin{equation}
g_a = e^{i \epsilon \sigma_a }
\end{equation}
on the group element $g$ of $SU(2)$ written in terms of canonical coordinates described above. 
Explicitly then we have that,
\begin{eqnarray}
g_{11}& = &  \cos (\psi) + i \sin(\psi) \cos(\theta) \\ \nonumber
g_{12}& = & i \sin(\psi) \sin(\theta) e^{-i \phi} \\ \nonumber
g_{21}& = & i \sin(\psi) \sin(\theta) e^{i \phi}\\ \nonumber
g_{22}& = & \cos (\psi) -i \sin(\psi) \cos(\theta) 
\end{eqnarray}
Expanding out to linear order in $\epsilon $ in the $g_i$ and the reading of the infinitesimal variations 
$\delta \psi, \delta \theta, \delta \phi $ and by equating
\begin{equation}
\delta X^i = \epsilon k^j_a \partial_j X^i
\end{equation}
we can then read off the components of the killing vector $k^i_a$ for  each of the isometries induced by left action with 
group element  $g_a , a=1,2,3$ of $SU(2)$. This gives,
\begin{eqnarray}
k_1 &=& [ \sin(\theta) \cos(\phi), \sin(\phi) - \cot(\psi) \cos(\theta)\cos(\phi), -\cot(\theta) \cos(\phi) -\cot(\psi) \frac{\sin(\phi )}{\sin(\theta)} ] \\ \nonumber
k_2 &=& [ -\sin(\theta) \sin(\phi), \cos(\phi) - \cot(\psi) \cos(\theta)\sin(\phi), -\cot(\theta) \sin(\phi) -\cot(\psi) \frac{\cos(\phi )}{\sin(\theta)} ] \\ \nonumber
k_3 &=& [\cos(\theta), -\cot(\psi)\sin(\theta), -1]
\end{eqnarray}
One can check (its a bit tedious!) that the above killing vectors are orthonormal with respect to to the canonical metric on $S^3$
\begin{equation}
k^i_a g_{ij} k^j_b = \delta_{ab}
\end{equation}
Although, using the above results, it is \emph{possible} therefore to show the $\sigma$ independence of the Lagrangian for geodesic windings, 
even when working in canonical coordinates, the resulting expressions are incredibly complicated. 
If we now return to our coordinate independent description - armed with the knowledge that the Lagrangian density \emph{must} 
be $\sigma$ independent for geodesic windings - we find that we can in fact reproduce most of our results without reference to a specific coordinate system.

\n
Looking at the effective potential (setting all time derivatives to zero) we find,
\begin{equation}
V = a_0^2 T_1 \sqrt{ r^2 + a_0^{-2}R^2 W }
\end{equation}
and we now know that the second term is just the warped length 
squared of the wrapped string on the $S^3$. The tangent vector to a geodesic $x'^i$
(where a dash here refers to differentiation with respect to the affine parameter along the curve) can always be defined to have unit length ie $x'^i g_{ij}x'^j =1$. 
In our case the string wraps closed curve geodesics with different winding numbers in general. 
A nice example of closed curve geodesics on the $S^3$ are the integral curves whose tangent vectors are
the three Killing vectors $k^i_a, a=1,2,3$ that generate the $SU(2)$ isometry group of $S^3$ (see above). Thus it is natural to 
take the following ansatz for our static wrapped string
\begin{equation}\label{static}
 X^{'i}_G(\sigma) = \sum_a n_a x'^i_a = \sum_a n_a k^i_a(X(\sigma)) 
\end{equation}
where we have used the fact that the tangent vectors to the $SU(2)$ generated geodesics are just the Killing vectors
$k^i_a$ (and where the subscript G implies that the embedding ansatz satisfies (\ref{geod}) and (\ref{preserv})). 
The Killing vectors not only have unit length with respect to the canonical metric on the $S^3$ but they are in fact orthonormal.
Using this it is easy to see that $W = \sum_a n_a n_a$, which is a constant and so is consistent with our previous analysis.

\n
The above makes sense because it is known that flows of the Killing vector fields on $S^3 $ induced by 
rigid $SU(2)$ rotations are actually geodesics. So our wrapping ansatz would be guaranteed to be 'minimal' in the same sense that the 
linear ansatz of Iglesias for winding around the torus - because again the flows of the Killing vectors in that case are geodesic circles around the torus.

\n
Notice that in equation (\ref{static}) we have written the expression for $X^{'i}(\sigma)$ and not $X^i(\sigma)$. The latter can be obtained
in principle by integration. In the simple case where the Killing vectors are coordinate independent (e.g. Abelian isometries, shift isometries), 
integration directly gives us $X_G^i = X^i_0 + \sigma n_a k^i_a$ which we found in the case of flat sub-manifolds
discussed earlier. The $SU(2)$ Killing vectors however are not constant, so integration is non-trivial and will lead to
non-linear dependence on $\sigma $ in general.

\n
It is a general result that if a Killing vector field has constant length, then it will generate geodesics along the manifold. In our case 
we know that the $k^i_a(X)$ have unit length with respect to the canonical metric on the $S^3$. Therefore we can
be sure that equation (\ref{geod}) will be solved by $X'^i(\sigma) \sim k^i_a(X)$ on the three sphere according to the ansatz in (\ref{static}).
One can see this more clearly by taking the normalisation constraint $g_{ij}k_a^i k_b^j = \delta_{ab}$ and differentiating this with respect to
some arbitrary vector field $Y$. The resultant Killing equation gives us $\nabla_{k_a} k_b = 0$, which more generally implies
\begin{equation}
\nabla_{n_a k_a} n_b k_b = 0
\end{equation}
which is just the restatement that $n_ak_a^i$ is a geodesic.

\n
Since we are interested in the dynamics of such an embedding we now need to extend our solution to incorporate time dependence. This means that
we need to preserve the geodesics under time evolution. To ensure this, let us use the $SU(2)$ transformations, since these are isometries preserving
the canonical metric, and therefore we map geodesics to geodesics. The modified ansatz for the embedding functions should now read
\begin{eqnarray}
X'^i(\sigma, t) &=& \sum_a n_a k_a^i(X(\sigma,t)) \nonumber \\
\dot{X}^i (\sigma,t) &=& \sum_a \lambda_a k^i_a(X(\sigma, t))
\end{eqnarray}
where we have introduced new variables $\lambda_a \in \mathbb{R}$. The string modes will therefore be wrapped along any curve defined by $n_a k_a^i$, and this
curve will then evolve in time via the second geodesic equation (with tangent vector $\lambda_a k_a^i$).
This means that the functions in the action will take a simplified form
\begin{eqnarray}
W &\to& \sum_a n_a n_a \\
P &\to& \sum \lambda_a \lambda_a \nonumber
\end{eqnarray}
which are both constant, and $\lambda_a$ is related to the average speed of the string along the $S^3$.

\n
Now it $n_a$ and $\lambda_a$ are such that the $X'^i$ and $\dot{X}^i$ are parallel, then the dangerous $R^4$ terms will vanish from the action. If these
vectors are not parallel then there is no cancellation, which means that the wrapped geodesic must have a perpendicular component along the winding direction
satisfying $\dot{X}^i X'^j g_{ij}=0$.

\n
How does this affect the resultant action constructed in equation (\ref{generalaction})? 
We can define a unit vector pointing parallel to the winding direction in the usual manner. This therefore
allows us to split the velocity into components $\dot{X}^i = \dot{X}^i_{\parallel} + \dot{X}^i_{\perp}$. After some manipulation we find that the
Lagrangian simplifies to give
\begin{equation}
L = -{\cal T} a^2 \sqrt{(1-\dot{r}^2-a^{-2}R^2 \dot{X}^i_{\perp}g_{ij}\dot{X}^j_{\perp})(r^2+a^{-2}R^2 W)-a^{-2}R^2 r^2 \dot{X}^i_{\parallel}g_{ij}\dot{X}^j_{\parallel}}
\end{equation}
where the $R^4$ terms have canceled as already advertised. 
Physically this makes sense, since the perpendicular modes lie in the normal bundle and therefore
contribute to the transverse boost of the string, much like the velocity in the Minkowski directions. 
The net effect is an enhancement of the relativistic 'gamma' factor.

\n
In the simplest case where we neglect the transverse modes, we see that the resultant energy and momentum become
\begin{eqnarray}
E &=& 2\pi T_1 \frac{a_0^2 (r^2 + a_0^{-2} R^2 W)}{\sqrt{(1-\dot{r}^2)(r^2 + a_0^{-2}R^2 W)-a_0^{-2}R^2 r^2 P}} \nonumber \\
l &=& 2\pi T_1 \frac{R^2 r^2 \lambda_a n_a}{\sqrt{(1-\dot{r}^2)(r^2 + a_0^{-2}R^2 W)-a_0^{-2}R^2 r^2 P}}.
\end{eqnarray}
where we have defined $l$ in the same manner as Iglesias
\begin{equation}
l = \frac{\delta L}{\delta \dot{X}^i} X'^i.
\end{equation}
For the case where there is no velocity in the Minkowski direction, we can write the energy as a function of $l$ and minimise it to obtain 
\begin{equation}
r_{*}^4 = \frac{l^2}{a_0^4 (2\pi)^2 T_1^2}
\end{equation}
which is the generalisation of the $r\sim  \sqrt{l}$ dependence obtained in Iglesias. Moreover the velocity at the minimal radius is given by
\begin{equation}
\dot{X}^2_{\parallel} = \lambda_a \lambda_a = \frac{a_0^2}{R^2}.
\end{equation}
Generally we should consider the case where there is non-zero $\dot{X}^i_{\perp}$, we will see that solutions that minimise the energy whilst
having non-zero $l$ \emph{require} us to set $\dot{r}=\dot{X}^i_{\perp}=0$ - which is easily understood since non-zero velocity in these directions only
ever increases the energy through enhancement of the 'gamma' factor.

\n
In the general case we see that the above expressions can be written in the modified form
\begin{eqnarray}
E &=& 2\pi T_1\frac{(r^2+a_0^{-2}R^2 n_a n_a)}{\sqrt{(1-\dot{r}^2-a_0^{-2}R^2 \dot{X}^i_{\perp} g_{ij} \dot{X}^j_{\perp})(r^2+a_0^{-2}R^2n_bn_b)-a_0^{-2}R^2r^2 \lambda_b \lambda_b}} \nonumber \\
l &=& 2\pi T_1 \frac{R^2 r^2 \lambda_a n_a}{\sqrt{(1-\dot{r}^2-a_0^{-2}R^2 \dot{X}^i_{\perp} g_{ij} \dot{X}^j_{\perp})(r^2+a_0^{-2}R^2n_bn_b)-a_0^{-2}R^2r^2 \lambda_b \lambda_b}}
\end{eqnarray}
%%%%%%%%%%%%%%%%%%%%%%%%%%%%%%%%%%%%%%% 

\end{document}